\documentclass[prX,showpacs,superscriptaddress,twocolumn]{revtex4}

\usepackage{graphicx}
\usepackage{amsmath}
\usepackage{amsfonts}
\usepackage{amssymb}
\usepackage{epsf}
\usepackage{hyperref}

\begin{document}
\title{A Maxwell's demon in the generation of an intense and slow guided beam}
\author{G. Reinaudi}
\affiliation{Laboratoire Kastler Brossel, Ecole Normale
Sup\'erieure, 24 rue Lhomond, 75005 Paris, France}
\author{D. Gu\'ery-Odelin}
\affiliation{Laboratoire Kastler Brossel, Ecole Normale
Sup\'erieure, 24 rue Lhomond, 75005 Paris, France}
\affiliation{Laboratoire Collisions Agr\'egats R\'eactivit\'e,
CNRS UMR 5589, IRSAMC, Universit\'e Paul Sabatier, 118 Route de
Narbonne, 31062 Toulouse CEDEX 4, France}

 \date{\today}

\begin{abstract}
We analyze quantitatively the generation of a continuous beam of
atoms by the periodic injection of individual packets in a guide,
followed by their overlapping. We show that slowing the packets
using a moving mirror before their overlapping enables an optimal
gain on the phase space density of the generated beam. This is
interpreted as a Maxwell's demon type strategy as the
experimentalist exploits the information on the position and
velocity of the center of mass of each packet.
\end{abstract}

\pacs{32.80.Pj, 03.75.Pp}

\maketitle

Over the last thirty years, there has been very significant and
impressive progress in the experimental ability to increase the
phase space density of atomic clouds, enabling the quantum
degenerate regime to be reached \cite{NobelBEC}.

All these advances can be revisited in terms of information
entropy \cite{KeP92}. The powerful laser cooling technique
\cite{Nobel98} decreases dramatically the temperature and the
entropy of an atomic cloud, at the expense of an increased
disorder for the photons leaving the laser mode through
spontaneous emission to populate other modes. The entropy of the
global system made of $\{$atoms+photons$\}$ increases as expected
from the second law of thermodynamics \cite{EnN92}. In the
evaporative cooling technique \cite{Hes86}, the disorder of the
system made of all particles involved from the beginning of the
evaporation ramp increases each time a particle is evaporated,
since this atom is no more localized in the trapping region.
Accordingly, this technique yields a decrease of entropy for the
subsystem made of the remaining trapped particles. Some of those
techniques have also been implemented on atomic beams
\cite{LWR05}.

\begin{figure}[b]
\centerline{\includegraphics[width=\columnwidth]{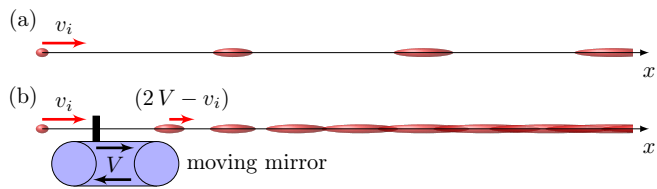}}
 \caption{(Color online) Schematic representation of the generation of a continuous
  beam by injecting packets into a guide (a), and by slowing them with a moving mirror (velocity $V$) before their overlapping (b).} \label{fig1}
\end{figure}

An optimized scheme for the implementation of informational
cooling has been recently proposed in Ref.~\cite{RMR06} and
experimentally demonstrated in Ref.~\cite{PBV08}: in this scheme,
the increase of entropy of the radiation field in the scattering
of a photon is exactly compensated by the  reduction of entropy
for the trapped atoms. Another strategy to increase the phase
space density consists in changing adiabatically, and therefore
isentropically, the density of states experienced by the atoms as
demonstrated in \cite{SMC98}. The gain in information results from
the transfer of population in the new set of low energy levels.
The entropy $S$ is simply related to the phase space density
$\rho$ by $S = - Nk_B\log \rho+S_0$ \footnote{According to the
canonical formalism in statistical physics, the one particle phase
space density is $\rho=N/z$, where $z = h^{-3}\int d^3rd^3p \exp
(- {\cal H}({\bf r},{\bf p})/k_BT)$ and ${\cal H}({\bf r},{\bf
p})$ is the one particle hamiltonian. The entropy $S$ is obtained
through the expression for the free energy $F=U-TS$, from which we
infer the value of the term $S_0/(Nk_B)=1+U/(Nk_BT)$. If the shape
of the trapping potential changes, the mean energy $U$ and
therefore $S_0$ change also.}. If the shape of the confining
potential is modified adiabatically, $S$ remains constant but
$S_0$ changes which modifies in turn the phase space density
$\rho$ \cite{PMW97,SMC98}.

Conversely, information can be used directly to increase the phase
space density. This is realized in the stochastic cooling
technique applied on a beam of charged particles in a storage ring
\cite{Mee85}. Taking advantage of the particle's charge,
information is extracted in one place and an adapted feedback
action in another place is exerted later on. This technique seems
at first sight to violate the Liouville theorem which states the
incompressibility of phase volume when only conservative forces
are involved. However particles being point like, there is a lot
of empty space between them. Each particle can in principle be
manipulated individually to increase the phase space density. This
requires all information about position and velocity of the
particles. Such a procedure resembles Maxwell's demon thought
experiment \cite{Max71}. There is no violation of the second law
of thermodynamics since the measurement performed by the demon
implies an entropy increase \cite{Szi29}.

It is definitely more difficult to extract information on a beam
made of neutral particles. We show in this article that a recently
published {\it optimization} \cite{RWC06} of the technique
presented in Ref.~\cite{LVG04} to generate a continuous beam by
periodically injecting packets of atoms in a guide is reminiscent
of Maxwell's demon strategy. The generation of an intense and slow
guided beam involves two conflicting requirements: the high flux
implies coupling packets at a high repetition rate, and the low
velocity requirement limits this rate. An upward potential hill
can be used to slow down the beam \cite{LWR05}. However, a better
strategy from the point of view of the phase space density of the
generated beam consists in slowing down the packets by letting
them undergo an elastic collision with a moving potential barrier
before their overlapping \cite{RWC06,NLR07} (see Fig.~\ref{fig1}).
The reason why this latter scheme can be better than the former
one in terms of entropy is that it corresponds to the realization
of a true Maxwell's demon with the use of information on the
center of mass of the packet before the overlapping.

\begin{figure}[b]
\centerline{\includegraphics[width=\columnwidth]{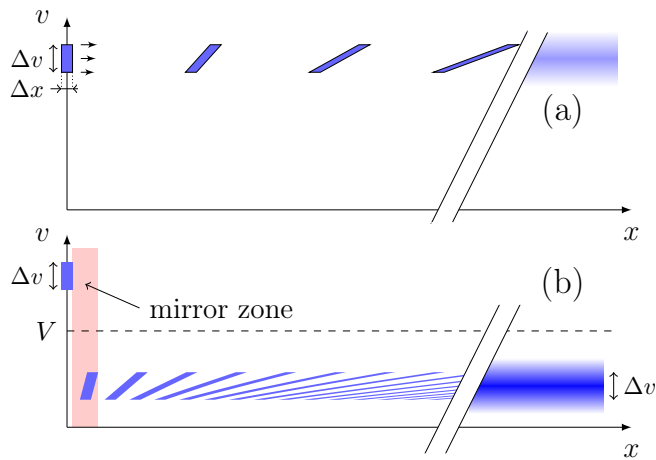}}
 \caption{(Color online) Representation in the single-particle phase space of one-dimensional atomic packets periodically injected in a guide.
 a) in the absence of slowing b) in the presence of slowing with the mirror moving at a velocity $V$ (dashed line). For a sufficiently
 long propagation distance, packets have merged and the thermal equilibrium for the corresponding beam is
 reached.} \label{fig2}
\end{figure}

For the sake of simplicity and without loss of generality, the
argument is presented quantitatively for a one-dimensional system.
However, all results derived in this paper are valid when one
takes into account explicitly the transverse degrees of freedom,
as long as the transverse confinement is not modified or only
modified adiabatically by the presence of the mirror, and the
initial velocity dispersion is the same on all degrees of freedom.

We assume the successive packets to be identical and uniformly
distributed over a rectangular surface in the one-dimensional
phase-space. The packets are then characterized by four
parameters: the number of atoms $N$, the initial size $\Delta x$,
the velocity dispersion $\Delta v$ and the center of mass velocity
$v_i$. The phase space density of a given packet reads:
\begin{equation}
\rho_p = \frac{N\hbar}{m\Delta x\Delta v}, \label{eq1}
\end{equation}

where $m$ denotes the mass of the atoms. Figure \ref{fig2}(a)
represents a plot of $\rho_p$ in the one-particle phase space at
different times. As expected from the Liouville theorem, the
surface occupied by the packet remains constant. However its size
in position space increases with time as a result of the velocity
dispersion, or in other terms, velocity-position correlations are
produced in the course of the time evolution of the packets. We
consider such packets launched periodically with a constant time
separation $\tau$. After a sufficient duration, they overlap,
thermalize and form a continuous beam \cite{LVG04}. This beam in
thermodynamic equilibrium is characterized by an atomic density
$N/(v_i\tau)$ and the same velocity dispersion $\Delta v$ as the
one of the packets \footnote{Strictly speaking, the thermalization
cannot occur in a one-dimensional system. We assume that
thermalization occurs which requires the problem to be three (or
two) dimensional. The effective one dimensional treatment carried
out in this article is valid if the longitudinal velocity
dispersion is unchanged after thermalization. It is the case if
the velocity dispersion along all degrees of freedom is initially
the same}. The phase space density of the thermalized beam is
given by:
\begin{equation}
\rho = \frac{N\hbar}{m v_i\tau \Delta v}.\label{eq2}
\end{equation}
The factor term $N/ v_i\tau$ corresponds to the mean atomic
density of the beam. From Fig. \ref{fig2}(a), we immediately
conclude that $\rho < \rho_p$. This inequality reflects the
Liouville theorem. It can be interpreted physically using the
concept of information entropy \cite{KeP92}. Before overlapping,
the packets are distinguishable ($v_i\tau > \Delta x$), therefore
the center of mass of each packet is well-defined. The overlapping
accompanied by elastic collisions between successive packets
corresponds to a loss of information on the center of mass of the
packets. This merging process yields an increase of entropy or
equivalently a decrease of the mean phase space density of the
beam generated from the packets compared to the one of each
packet.

An important feature of Eq.~(\ref{eq2}) is that the mean velocity
of the packet enters explicitly the expression of the phase space
density of the beam. In Ref.~\cite{RWC06}, this velocity
dependence was exploited to realize a slow and intense guided
atomic beam. Each atomic packet was slowed down by means of a
moving mirror well-synchronized with the motion of the atomic
packet. The overlapping occurred after this manipulation of each
packet. This experimental trick permits one to ensure a high flux
while having, in the end, a very low mean velocity for the beam
generated from the slowed packets.

Such a specific action on each packet is reminiscent of the
Maxwell's demon thought experiment. In Maxwell's scheme, the
apparent violation of the second law of thermodynamics is made by
exploiting information about particle's velocity. The
experimentalist acts as a Maxwell's demon by exploiting all the
information (position and velocity) on the center of mass of each
packet to synchronize the motion of a moving mirror with which
they will undergo an elastic collision\footnote{The system made of
the succession of the packets can be regarded as the one of a
succession of macro-particles, each macro-particle being a packet.
The action of the Maxwell's demon in this context consists in
increasing the linear density of the macro-particles by slowing
them down one after another.}. The macroscopic mirror absorbs the
microscopic momentum kick due to the reflection of the packet in
the mirror's frame.

 The
reflection of a succession of slowed packets is represented on
Fig. \ref{fig2}(b). In the single-particle phase space, it
corresponds essentially to a translation while keeping its volume
constant. The phase space density of the beam generated from those
slowed packets is given by
\begin{equation}
\rho' = \frac{N\hbar}{m(2V - v_i)\tau \Delta v} = \rho
\frac{v_i}{(2V - v_i)}> \rho,\label{eq3}
\end{equation}
where $(2V-v_i)$ represents the mean velocity of the beam made of
the packets that have been slowed down through their interaction
with the moving mirror. For the same repeating rate $\tau^{-1}$, a
significant gain on the phase space density of the generated beam
is therefore achievable.

The upper bound on phase space density of the beam generated from
packets is given by the phase space density of each packet: $\rho'
< \rho_p$. This holds even if one uses time dependent potentials
to manipulate them before their overlapping. In this context,
there is no possible strategy to overcome this limit \cite{KeP92},
since we do not use for each packet information at a microscopic
scale i.\,e. on atoms individually.

For a given set of packet parameters $(N,v_i,\Delta v,\Delta x)$,
one may wonder what is the optimum choice of mirror velocity $V$
to maximize the phase space density $\rho'$ of the beam generated
from the slowed packets, and what is the expression for this
optimum depending on the experimental parameters.

To answer those questions, we will consider a specific example
which contains all the relevant physical ingredients. As mentioned
above, we model the initial packet by a uniform phase space
density with an initial rectangular shape. The coordinates of the
four vertices are $(\Delta x/2,v_i+\Delta v/2)$, $(-\Delta
x/2,v_i+\Delta v/2)$, $(\Delta x/2,v_i-\Delta v/2)$ and $(-\Delta
x/2,v_i-\Delta v/2)$. The mirror velocity $V$ has two constraints:
the lowest velocity of the initial packet needs to be larger than
the mirror velocity $v_i-\Delta v/2
>V$, and the lowest final velocity after interaction with the
mirror should be positive to ensure the propagation of the beam in
a well-defined direction $2V - v_i-\Delta v/2>0$. Let us introduce
the dimensionless parameters $y=\Delta v/v_i$ and $z=V/v_i$ that
are related respectively to the packet and the mirror. The two
previous conditions on the mirror velocity are recast in the form:
\begin{equation}
\frac{1}{2}+\frac{y}{4} \le z \le 1- \frac{y}{2}.\label{eq4}
\end{equation}

To maximize the flux of the continuous beam resulting from the
overlapping of the periodically injected atomic packets, one needs
to maximally decrease the time period $\tau$ between two
successive packets \cite{RWC06}. However, $\tau$ cannot be chosen
arbitrary small since the mirror is moving faster than the slowed
packets and should not push and thus accelerate some of them while
interacting with the next packet. In the following, we denote
$\tau_{\rm min}$ the minimum repeating time that enables the
mirror to slow down a given packet while not affecting the
preceding slowed packet. This quantity depends on $(\Delta
x/v_i)$, $y$ and $z$. To work out its explicit expression, we
model the mirror by an infinitely high and thin potential barrier.
The mirror is periodically moving at a velocity $V$ over a
distance allowing the slowing of all atoms of each packet and we
assume that it acts on a given packet as soon as it is released
(see Fig.~\ref{fig2}b). We find:
\begin{equation}
    \tau_{\rm min}(y,z) = \frac{\Delta x}{v_i}\,{\frac {\left( z-1-y/2\right) }{\left( 2\,z-1-y/2
 \right)  \left( z-1+y/2\right) }}
    \label{eq:Tmin}
\end{equation}

\begin{figure}[b]
\centerline{\includegraphics[width=\columnwidth]{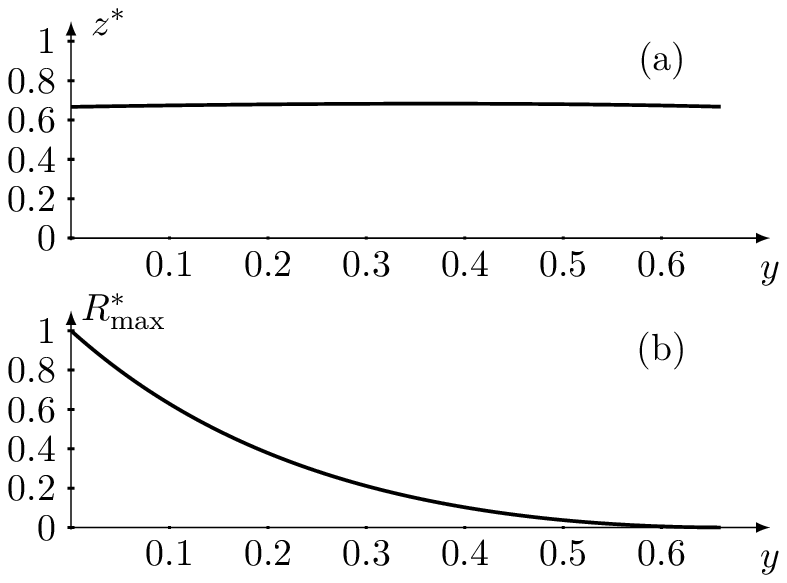}}
 \caption{
 (a) The
 optimal mirror velocity $z^*(y)=V^*/v_i$ normalized to $v_i$ is plotted as a function of the dimensionless parameter $y=\Delta v/v_i$ and (b) the corresponding
 maximum phase space density $R^*_{\rm max}=R_{\rm max}
(y,z^*(y))= \rho'/\rho_p$ of a beam generated from the packet
after
 their optimal slowing down and normalized to the packet initial phase space density, is plotted as a
 function of $y$.} \label{fig3}
\end{figure}

Using this result with Eqs.~(\ref{eq1}) and (\ref{eq3}), we infer
the maximum increase in phase space density:
\begin{eqnarray}
R_{\rm max} (y,z) = \frac{\rho'_{\rm max}}{\rho_p}  =
\frac{\Delta x}{(2V-v_i)\tau_{\rm min}(y,z)}.\label{eq5}
\end{eqnarray}

For given experiment, the dimensionless parameter $y$ is fixed.
The optimum value of the mirror velocity $V^*$ is obtained by
maximizing $R_{\rm max} (y,z)$ as a function of $z$. Taking into
account the constraints (\ref{eq4}), the equation $\partial R_{\rm
max} (y,z)/\partial z=0$ gives a unique solution $z^*(y)$:
\begin{equation}
V^*= v_i z^*(y) = \frac{v_i}{6} \left( 2+y +\sqrt{2}\sqrt{2-y-y^2}
\right),\label{eq6}
\end{equation}
and the domain in $y$ for which a solution exists is: $y < 2/3$.
This condition simply means that the initial mean velocity has to
be large enough compared to the velocity dispersion, as
intuitively expected.

As illustrated on Fig.~\ref{eq3}(a) by plotting $z^*(y)$ from
Eq.~(\ref{eq6}), a remarkable feature of the optimal velocity for
the mirror is that it is nearly constant over its validity domain
and approximately equal to $2v_i/3$. We conclude from
Eq.~\ref{eq3} that an optimal use of the mirror technique permits
to gain a factor on the order of three on the phase space density
generated from the packets compared to the value obtained in the
same conditions but in the absence of the mirror.

Figure \ref{fig3}(b) shows that when $\Delta v/v_i$ tends to zero,
the phase space density of the continuous flow tends to its upper
bound, i.e. the phase space density of the individual packets.
This corresponds to a situation where the slowed packets cover the
single atom phase space in a quasi compact manner.

In practice, two effects tends to reduce the gain on the phase
space density of a beam generated from slowed packet compared to
the one without slowing \cite{RWC06}: (i) the finite thickness of
the mirror, and (ii) a free propagation of the packets before
their interaction with the mirror. We evaluate separately their
effect in the following.

We denote $\Delta_{\rm m}$ the thickness of the mirror. For
example, $\Delta_{\rm m}\sim 10$ cm in the experiment described in
 Ref.~\cite{RWC06}. The
calculations performed previously can be readily adapted to take
into account the size of the mirror. In the limit $y\rightarrow
0$, the maximum gain on phase space density $R'_{\rm
max}(y,z^*(y))$ saturates to $1/(1 + \Delta_{\rm m}/\Delta x)$.
Indeed, the incompressible distance $\Delta_{\rm m}$ dictates an
upper limit on the achievable atomic density. In addition, the
optimal mirror velocity $V^*$ tends to $v_i$ as $\Delta_{\rm m}$
increases, which reflects the reduction of the gain on phase space
density resulting from the limit on the atomic density.

Another experimental parameter to be considered lies in the fact
that the atomic cloud cannot usually be slowed down just after its
injection, but has to propagates freely over a distance $D$ before
interacting with the mirror \footnote{Indeed, a finite time for
the preparation of a packet is required and one should ensure that
the preparation of a given packet does not affect the previous
packets, for instance through scattered light from the
magneto-optical trap used to prepare the packets \cite{LVG04}.}.
For example, one has $D \sim 25$ cm in Ref.~\cite{RWC06}. In this
instance and assuming that $ \Delta_{\rm m}=0$ for sake of
simplicity, the general expression for the maximum of the ratio
$\rho'/\rho_p$ takes the form:
\begin{equation}
R'_{\rm max} (y,z,D)  = \frac{ R_{\rm max} (y,z)}{1+y\,\dfrac{
D}{\Delta x}}. \label{eq9}
\end{equation}
This result just reflects the fact that the packets have spread
during their free propagation before interacting with the mirror,
which reduces as expected the gain compared to the one without
free flight. From Eq.~(\ref{eq9}), we conclude that the optimal
velocity $V^*$ is the same as the one calculated Eq.~(\ref{eq6}).

In conclusion, we have investigated quantitatively an optimal
strategy to produce a continuous beam with individual packets, in
order to maximize the phase space density of the beam. The use of
a mirror to slow down the packets before their overlapping can be
interpreted as a Maxwell's demon type strategy, where the mirror
acts as an active valve that modifies the properties of the packet
by reducing its mean velocity. This study exemplifies, in the
context of an atomic beam made of neutral atoms, the usefulness of
the link between information, entropy and phase-space densities
for designing optimal strategies.

We are indebted to J. Dalibard, T. Lahaye, M. Jeppesen, R.
Mathevet, G. L. Gattobigio and A. Ridinger for useful comments.
We acknowledge 
 financial support from the D\'el\'egation G\'en\'erale pour
l'Armement (contract number 05-251487), the Institut Francilien de
Recherche sur les Atomes Froids (IFRAF) and the Plan-Pluri
Formation (PPF). G.~R. acknowledges support from the
D\'el\'egation G\'en\'erale pour l'Armement.

\end{document}